\documentclass[aps,prx,superscriptaddress,reprint]{revtex4-2}

\usepackage{graphicx}
\usepackage{lineno}
\usepackage{dcolumn} 
\usepackage{bm}
\usepackage{textcomp}
\usepackage{textcomp}
\newcommand{\nocontentsline}[3]{}
\newcommand{\toclesslab}[3]
{\vspace{0.1in}\bgroup\let\addcontentsline=\nocontentsline#1{#2\label{#3}}\egroup\vspace{-0.1in}}
\usepackage{xcolor}
\usepackage{soul}
\usepackage{newfloat}
\DeclareFloatingEnvironment[
  fileext=lof,
  name=Extended Data Fig.,
  listname=List of Extended Data Figures
]{extendedfigure}
\usepackage[%
  colorlinks=true,
  urlcolor=blue,
  linkcolor=blue,
  citecolor=blue
]{hyperref}
\begin{document}

\title{
Reversibility, Chaos, and Attractors in Periodically Sheared Elastic Filaments}

\author{Francesco Bonacci}
\altaffiliation[Equal contribution]{}
\affiliation{Department of Physics and Geology, University of Perugia, Via A. Pascoli, Perugia, 06123, Italy}

\author{Brato Chakrabarti}
\altaffiliation[Equal contribution]{}
\affiliation{International Centre for Theoretical Sciences, Tata Institute of Fundamental Research, Bengaluru 560089, India}

\author{Olivia du Roure}
\affiliation{PMMH-ESPCI, Université PSL, CNRS, Sorbonne Université, Université Paris Cité, 75005 Paris, France}

\author{Anke Lindner}
\affiliation{PMMH-ESPCI, Université PSL, CNRS, Sorbonne Université, Université Paris Cité, 75005 Paris, France}

\author{David Saintillan}
\affiliation{Department of Mechanical and Aerospace Engineering, University of California San Diego, La Jolla, California 92093, USA}

\begin{abstract}
The dynamics of filaments in flow are central to understanding a wide range of biological and soft-matter systems, yet their behavior under time-dependent forcing remains poorly understood. Here, we investigate the long-time dynamics of Brownian inextensible elastic filaments subjected to strong uniform oscillatory shear by combining microfluidic experiments on actin filaments with numerical simulations based on a fluctuating Euler-Bernoulli elastica model in a viscous fluid. As the oscillation period increases, irreversibility emerges from the interplay of flow-induced deformations and thermal noise. This leads to a departure from reversible, deterministic rigid-body dynamics: in this regime, the filaments cycle between nearly straight, flow-aligned conformations at full periods and buckled shapes at half periods. Owing to the time-glide symmetry of the system, two such attracting states in fact coexist with a phase shift of half a period. The system spontaneously selects one, but occasionally switches between them as a result of noise, producing intermittent transitions between apparent order and disorder. This system constitutes an experimentally accessible realization of stochastic symmetry breaking, attractor hopping, and intermittency in a minimal nonequilibrium soft-matter system, with novel implications for the design and control of soft matter systems under time-dependent flows.

\end{abstract}

\maketitle

\section{Introduction}

Nonlinear systems driven by oscillatory forcing, ranging from sheared soft materials \cite{PGBL2005,GJA2020,KP2021,YZXYZKW2024,CCCF2024, Galloway2022} and biological networks \cite{FP2003,FKTJDMTSLBVDR2014} to quantum systems \cite{GPSZ2006,GD2014}, exhibit rich dynamical behaviors, including transitions from reversible, periodic motion to chaotic, irreversible states \cite{RRDOR2023}. In these so-called Floquet systems, reversibility is typically assessed stroboscopically, by comparing the system’s state at the beginning and end of each driving period $T$. Reversible to irreversible transitions occur across many interacting systems: under certain parameter values or driving frequencies, the system self-organizes into a reversible state where microscopic trajectories repeat during each cycle, while for other parameter values, trajectories exhibit chaos with no clear periodicity. 

Periodically sheared viscous materials have served as model systems to study such transitions. In the absence of Brownian motion and at low Reynolds number, the linearity of the Stokes equations guarantees kinematic reversibility for isolated rigid particles under time-reversal of the flow \cite{GM2011}. However, when multiple particles interact, reversibility can break down due to interparticle collisions. Pine \textit{et al.} \cite{PGBL2005} famously demonstrated a sharp transition at a critical strain in colloidal suspensions. Below the critical strain, particles are found to rearrange in the first few cycles of driving as a result of collisions, reaching a reversible ``absorbing'' state in which all collisions are avoided. Above it, collisions continue occurring over arbitrarily many cycles, leading to irreversible particle diffusion. Related phenomena have been observed in driven emulsions \cite{JB2014}. In both particle \cite{WGPC2020} and droplet \cite{WJDB2015} systems, the emergent reversible state typically exhibits spatial hyperuniformity \cite{T2018}. Brownian fluctuations can blur the transition, disrupting reversibility even at low strain.

 
Beyond interactions, particle deformability introduces another route to irreversibility, as exemplified by the case of a semiflexible elastic filament. In shear \cite{BS2001,TS2004,HWPKB2013,LCDLD2018,SXSNSE2022} as well as strain-dominated \cite{YS2007,KG2012,GKSS2012,MS2015,QSDL2015,CLLCFDSL2020} flows, viscous stresses can overcome elastic resistance and induce buckling instabilities, whose onset and morphologies are governed by the filament’s elastoviscous number \cite{DLNS2019}. These instabilities are highly  sensitive to initial conditions and typically require thermal noise for activation:   in the absence of fluctuations, an isolated elastic filament will simply relax to a straight, rod-like configuration and follow reversible dynamics \cite{TS2004}. This suggests a binary outcome for filament dynamics under oscillatory shear: either reversible straight motion (without noise) or fully irreversible deformation (with noise).


In this work, we reveal a subtler and previously unreported regime. Even in the presence of Brownian fluctuations, we report the emergence of long-lived, quasi-reversible states in which filaments undergo dramatic shape changes within a cycle but repeatedly return to nearly straight conformations upon flow reversal. Two such attractors in fact coexist due to the symmetry of the system, giving rise to intermittent transitions between apparent order and disorder. 

We combine microfluidic experiments on actin filaments, Brownian dynamics simulations, and a reduced-order normal mode model to explore this behavior across a range of driving frequencies. Our analysis focuses on the regime of strong flows, where buckling instabilities are prevalent, and reveals a strong dependence on the dimensionless period of oscillation $\rho=\dot{\gamma}_m T$, where $\dot{\gamma}_m$ is the maximum applied shear rate.  At low $\rho$, oscillations are too rapid to induce significant deformation, and the filament exhibits nearly rigid, reversible dynamics. As $\rho$ increases, we identify a transition to a regime of intermittent chaos punctuated by stroboscopically reversible configurations. Our findings also highlight how shape instabilities can act as an effective noise source, driving complex behavior even in weakly fluctuating systems.

\section{Problem definition}

\begin{figure*}[t]
  \centering
  \includegraphics[width=\textwidth]{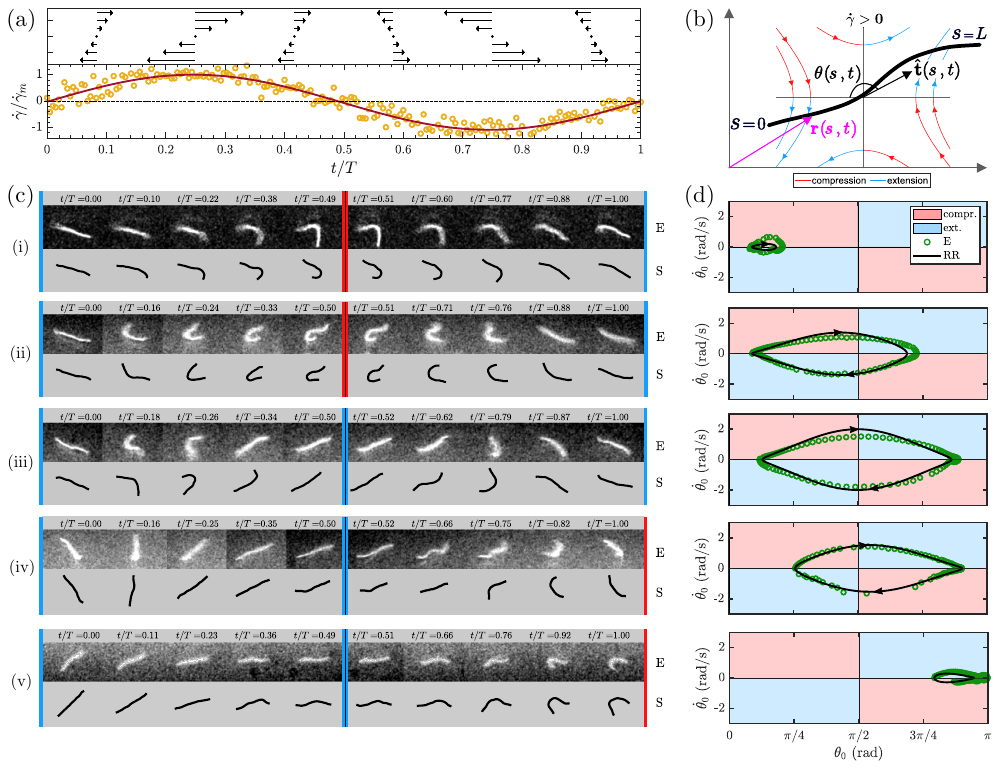}\vspace{-0.5cm}
  \caption{(a) Evolution of the time-periodic shear flow, with shear rate $\dot{\gamma}(t)/\dot{\gamma}_m =\sin{(2\pi t/T)}$, over the course of one period $T$ of oscillation. The plot compares experimental microfluidic shear rate measurements with a sinusoidal wave. (b) Schematic of a filament with arclength $s$, whose shape is described by the tangent vector $\hat{\mathbf{t}}(s,t)$, or equivalently by the tangent angle $\theta(s,t)$ measured from the negative $x$-direction. The positive $x$-direction is defined as the flow direction at $t=0$. The background flow lines illustrate the underlying straining flow, which is extensional in the first and third quadrants and compressional in the second and fourth quadrants during the first half period when $\dot{\gamma}>0$. (c) Snapshots of representative filament trajectories over the course of one period, from experiments (E) and simulations (S) with $\ell_p/L = 1.3$, $\bar{\mu}_m \approx 1 \times 10^4$. The dimensionless period is $\rho=6$ in (i) and $\rho=13$ in (ii)-(v). Cases (ii)-(v) differ by the initial orientation of the filament. The blue and red vertical lines label the deformation state of the filament at the start and end of each half cycle: straight (blue) or deformed (red). (d) Orientational dynamics in the ($\smash{\theta_0,\dot{\theta}_0}$) phase plane, for the same five cases as in (c). In each case, the experimental trajectory (E) is compared to the deterministic non-Brownian rigid-rod trajectory (RR) governed by Jeffery's equation Eq.~(\ref{eq:Jeffery}). Deviations from Jeffery orbits coincide with periods of strong filament deformation.}\vspace{-0.0cm}
  \label{fig:1}
\end{figure*}

We investigate the long-time behavior of isolated, Brownian, inextensible elastic filaments subjected to a uniform oscillatory shear flow with imposed shear rate $\dot{\gamma}(t) = \dot{\gamma}_m \sin{(2\pi t/T)}$, where $\dot{\gamma}_m$ is the peak shear rate and $T$ is the oscillation period [Fig.~\ref{fig:1}(a,b)].\ We combine microfluidic experiments using stabilized F-actin filaments with numerical simulations based on a fluctuating Euler-Bernoulli elastica model (see Appendix~\ref{appendixA} and \ref{appendixB} and \cite{bonacci2023dynamics} for details). The filaments are characterized by their length $L$, aspect ratio $\varepsilon$, persistence length $\ell_p$ and bending rigidity $B=k_B T\ell_p$ with $k_B T$ the thermal energy.

In the absence of inertia (and thus at vanishing Reynolds number), the filament dynamics are governed by three key dimensionless parameters:\ (i) the elastoviscous number $\bar{\mu}_m=8\pi\mu \dot{\gamma}_m L^4/Bc$ comparing viscous and elastic stresses \cite{DLNS2019}, where $\mu$ is the fluid viscosity and $c=-\log(e\varepsilon^2)>0$ is a geometric slenderness parameter, (ii) the dimensionless oscillation period $\rho=\dot{\gamma}_m T$, which also quantifies the accumulated strain over one half-cycle, and (iii) the ratio $\ell_p/L$, characterizing the strength of thermal shape fluctuations. The above dimensionless numbers are obtained by scaling lengths with $L$, time with $\dot{\gamma}_m$, and forces by $B/L^2$. In steady shear, buckling instabilities arise when $\bar{\mu}_m\gtrsim 306.4$ \cite{BS2001,LCDLD2018}.  

We examine the dynamics and morphology of the flexible filaments while they rotate back and forth through the compressive and extensional quadrants of the shear flow. All experiments and simulations presented here are conducted at a constant value of $\bar{\mu}_m$ much larger than the buckling threshold. Filament buckling can thus occur in the compressive part of the flow, and filaments are stretched in the extensional quadrant. Our analysis focuses on the influence of the driving frequency, captured by $\rho$, on the long-time behavior of the filaments.

Under shear, the filaments predominantly remain in the flow plane, allowing a two-dimensional description of their configuration \cite{LCDLD2018}. The instantaneous filament shape is represented by the tangent angle $\theta(s,t)$ to the centerline, where $s\in[0,L]$ is arclength and $\theta\in[0,\pi]$ is measured from the negative $x$-axis defined as opposite to the flow direction at $t=0$ [Fig.~\ref{fig:1}(b)]. To characterize the filament orientation and shape, we decompose $\theta(s,t)$ into normal modes: 
\begin{equation}
\theta(s,t)=\theta_0(t)+\sum_{k=1}^\infty \psi_k(t)\xi^k(s),
\end{equation}
where the mode shapes $\xi^k(s)$ are normalized eigenfunctions of the biharmonic operator \cite{Munk2006} (see Supplemental Material and Fig. S1 \cite{Supplemental_Info} for details). This provides a mode amplitude vector,
\begin{equation} \label{phasespace_vector}
\mathbf{\theta}(s,t)\to [\theta_0(t),\psi_1(t),\psi_2(t),...\psi_N(t)] = [\theta_0(t),\mathbf{\Psi}(t)], 
\end{equation}
which is truncated at an appropriate order $N$ and fully defines the filament's state at time $t$.\ The zeroth-order mode $\smash{\theta_0(t)=L^{-1}\int_0^L \theta(s,t)\,\mathrm{d}s}$ gives the average orientation, while higher-order modes $\mathbf{\Psi}(t) = [\psi_1(t),\psi_2(t),...\psi_N(t)]$ quantify shape deformations at increasing spatial frequencies.\ A global measure of the degree of deformation of the filament is provided by the norm $\|\mathbf{\Psi}\|$. In practice, we find that a small number of modes is sufficient to provide a good approximation to the filament shape (see SM and Fig. S1 \cite{Supplemental_Info}), and we use $N=5$ in the results below.

The imposed shear flow, illustrated in Fig.~\ref{fig:1}(a), displays a positive shear rate and clockwise vorticity in the first half of every period, followed by a negative shear rate and counterclockwise vorticity in the second half. As a result, a filament in this flow undergoes alternating half periods of clockwise  ($\dot{\theta}_0>0$) and counterclockwise  ($\smash{\dot{\theta}_0<0}$) rotation. In the absence of deformation ($\mathbf{\Psi}=\mathbf{0}$) and thermal noise, the angular dynamics are deterministic and reversible and follow Jeffery's equation for a rigid rod (RR) in planar shear \cite{Jeffery1922},
\begin{equation}
\frac{\mathrm{d}\theta_0^J}{\mathrm{d}t}=\sin^2\!\left(\theta_0^J\right)\, \sin (2\pi t/\rho),  \label{eq:Jeffery}
\end{equation}
where time has been non-dimensionalized using the typical flow time $\dot{\gamma}_m^{-1}$. Eq.~(\ref{eq:Jeffery}) provides a simple baseline for analyzing the effect of deformations and fluctuations in our observations below. In a time-independent shear flow, the angular velocity reaches a maximum when the filament is aligned with the flow-gradient direction ($\theta_0\approx \pi/2$) and vanishes when it is aligned with the flow axis ($\theta_0\approx 0$ or $\pi$). As shown in Eq.~(\ref{eq:Jeffery}), in oscillatory flow, the angular velocity remains a function of the filament's instantaneous orientation but is also modulated in time. The angular velocity is zero at the reversal of the flow direction every half period ($t=0$ and $t=T/2$), and the filament rotation slows down more gradually for reversals occurring close to the flow axis ($\theta_0\approx 0$ or $\pi$). 
Depending on its instantaneous orientation in the flow and on the sign of the shear rate, the filament either experiences compression or stretching due to the underlying straining flow, as depicted in Fig.~\ref{fig:1}(b) during the first half period when $\dot{\gamma}>0$. Under sufficient compression, buckling can occur \cite{BS2001,LCDLD2018,bonacci2023dynamics}. 

The angular dynamics of a Brownian rigid rod (BRR) are still described by Eq.~(\ref{eq:Jeffery}), with the right-hand side augmented by a stochastic forcing term $\xi(t)$, with variance $\langle \xi(t) \xi(t')\rangle = 4 \delta(t-t')/\mathrm{Pe}$,  where the P\'eclet number $\mathrm{Pe}= \dot{\gamma}_m/d_r$ compares the shear rate magnitude to the rotational diffusivity $d_r$. For the Brownian flexible filaments of primary interest here, thermal fluctuations are captured by the ratio ${\ell_p}/{L}$ characterizing the strength of thermal shape fluctuations. Note that these fluctuations not only lead to deviations from a straight shape but also induce noise in the mean orientation of the filament, equivalent to rotational diffusion.   Overall, deformations due to buckling, together with thermal fluctuations, introduce departures from the deterministic Jeffery dynamics  (see Fig.~\ref{fig:1}) and can lead to irreversibility in the long-time behavior. 


A key feature of the stochastic system studied here is its invariance under the transformation $t\rightarrow t+T/2$, $\theta\rightarrow \pi-\theta$, a property known as time-glide symmetry \cite{MPV2017}.\ This symmetry implies that the statistical behavior of the system over a full cycle remains unchanged under a half-period time shift combined with angular reflection. As we will show, this symmetry plays a key role in shaping the long-time asymptotic dynamics.   

\section{Results and discussion}

\subsection{Dynamics over one oscillation period} \label{sec:1-period}

We begin by characterizing the range of dynamical behaviors that arise over a single oscillation period, driven by the interplay between thermal fluctuations and buckling-induced deformations. The half-period dynamics of initially straight filaments were explored in our previous work \cite{bonacci2023dynamics}, which showed that both the oscillation period $\rho$ and the initial orientation $\theta_0^i = \theta_0(0)$ strongly influence the potential for buckling and the subsequent conformational dynamics. When multiple half periods occur in succession, both the filament orientation $\theta_0$ and deformation state $\mathbf{\Psi}$ at the start of each half cycle reflect a complex history of prior rotations and buckling events. In particular, a filament may begin a half cycle already deformed if it previously buckled and did not fully relax. These observations highlight the presence of dynamical memory in the system: the filament's response in a given cycle depends on its shape and orientation inherited from previous oscillations.


Figure~\ref{fig:1}(c) shows representative filament shapes from both experiments and simulations that illustrate common scenarios over one full period. The observations confirm the good agreement between experiments and simulations, which we have previously demonstrated to be quantitative \cite{bonacci2023dynamics}.

All examples shown correspond to cases where the filament is nearly straight ($\|\mathbf{\Psi}\|\approx 0$) at $t=0$, while the initial orientation is either contained in the compressional (cases (i)-(iv), $\theta_0^i\in[0,\pi/2]$) or extensional quadrant (case (v), $\theta_0^i\in[\pi/2, \pi]$) of the flow. While all examples exhibit transient deformation during the cycle, the timing and evolution of these deformations differ markedly. In cases (i) and (ii), the filament buckles during the first half cycle then gets entirely stretched in the second half.\ In case (iii), buckling then stretching occur in both half cycles in a nearly symmetric fashion.\ In case (iv), the filament simply tumbles without buckling during the first half period, but undergoes buckling in the second. In case (v), dynamics are comparable to (iv), despite the filament being initially oriented directly in the extensional quadrant.

We relate these behaviors to the evolution of the mean orientation angle $\theta_0$ and its angular velocity $\dot{\theta}_0$ in the phase portraits of Fig.~\ref{fig:1}(d), which compare experimental data to a numerical solution of Eq. (\ref{eq:Jeffery}) for a rigid non-Brownian rod. 
In case (i), the initial orientation is nearly aligned with the flow axis and the driving period $\rho$ is low. As a result, the filament remains in the compressional quadrant for the first half cycle ($\dot\theta_0>0$) and in the extensional quadrant for the second ($\smash{\dot\theta_0<0}$). After the initiation of buckling, the deformation therefore keeps growing until 
$t=T/2$, and subsequently relaxes when the flow is reversed. In case (ii), where $\rho$ is higher, the filament enters  the extensional quadrant before $t = T/2$, leading to partial stretching before reversal. As $\theta_0^i$ increases in cases (iii) and (iv), the filament enters the extensional quadrant earlier during the first half cycle, resulting in greater relaxation at $t = T/2$. Case (iii) shows a nearly symmetric trajectory in phase space, with alternating compression and extension in both half periods and clear buckling–relaxation cycles. In case (iv), the initial orientation is close to $\pi/2$, so the time under compression in the first half cycle is too short for any appreciable deformation to grow beyond the noise level set by thermal fluctuations \cite{bonacci2023dynamics}; as a result, buckling is not observed and the dynamics is akin to a rigid Brownian rod. In the second half cycle, however, the filament undergoes prolonged compression, resulting in buckling and a deformed shape at $t=T$. In case (v), the filament starts in the extensional quadrant and remains confined to orientations close to the flow direction, where angular velocities are small. As a result, the filament slowly approaches the flow direction during the first half cycle but never crosses it; buckling is observed in the second half cycle.

\begin{figure*}[t]
  \centering
  \includegraphics[width=\textwidth]{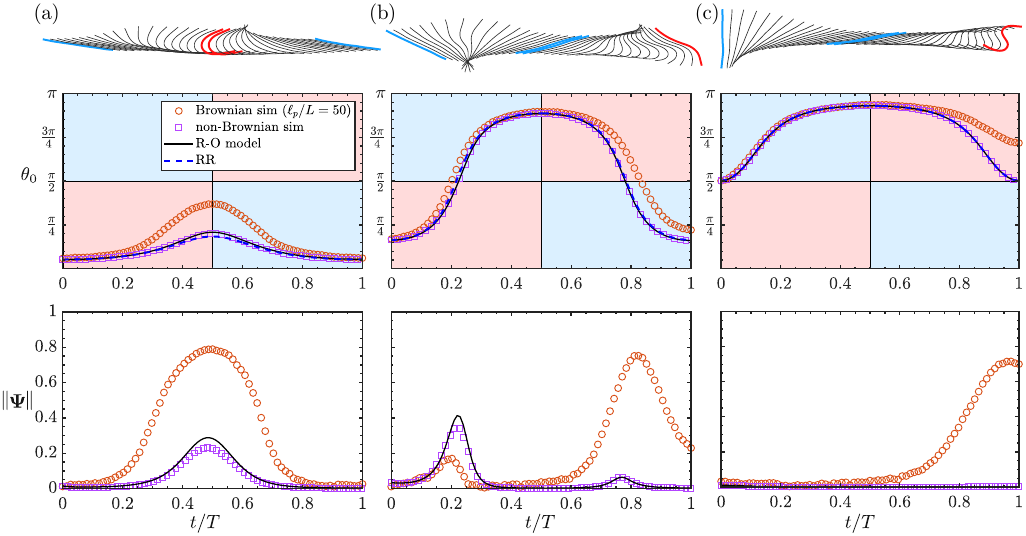}\vspace{-0.1cm}
  \caption{Mode dynamics from simulations and modeling over a single oscillation period for three filaments with different initial orientations: (a) $\theta_0^i = 10^{\circ}$, (b) $30^{\circ}$, and (c) $90^{\circ}$. Top row: Filament conformations from Brownian flexible simulations, where the conformations at the start and end of each half-cycle are highlighted using the same color code as in Fig.~\ref{fig:1}(c). Middle row: Time evolution of the mean filament orientation $\theta_0(t)$. Brownian simulation results (orange  circles) are compared with non-Brownian simulations (purple squares), the predictions from the semi-analytical reduced-order model (black solid line; see SM and Fig. S2 \cite{Supplemental_Info}), and the Jeffery orbit for a rigid rod (RR, blue dashed line). The colored background highlights regions of compressional and extensional flow, following the same color scheme as in Fig.~\ref{fig:1}(d). Bottom row: Time evolution of the filament deformation $\|\boldsymbol{\Psi}(t)\|$, computed using the first five normal modes. The legend matches that of the middle row. All cases use a dimensionless period $\rho = 14$ and an elastoviscous number $\bar{\mu}_m = 10^4$. Brownian simulations are performed with $\ell_p/L = 50$.}\vspace{-0.0cm}
  \label{fig:2}
\end{figure*}

Based on these observations, we identify some recurrent patterns in the dynamics that will be essential for understanding filament behavior on longer timescales. For this purpose, we assign a color label to the filament’s deformation state at the beginning and end of each half period: blue for a straight filament, and red for a deformed one (see vertical markers in Fig.~\ref{fig:1}(c)). This allows us to track sequences of deformation states over one full cycle. The most common sequences observed are: blue-red-blue (cases (i) and (ii)), red-blue-red (not explicitly illustrated in Fig.~\ref{fig:1}(c) but compatible with the cases shown under time-glide symmetry), and blue-blue-blue (case (iii)). Less frequent sequences include blue-blue-red (cases (iv) and (v)) and red-red-blue. The time-glide symmetry becomes obvious from these observations, visible by comparing cases (i) and (v), as well as (ii) and (iv).

Another striking feature from these examples is that the orientation angle $\theta_0$ at the end of the period often returns close to its initial value. This suggests that the mean angular dynamics approximately follow the deterministic evolution described by Jeffery’s equation, even in the presence of thermal fluctuations and filament deformability. Closer inspection of Fig.~\ref{fig:1}(d) shows that deviations from Jeffery orbits do in fact arise, and coincide with periods of significant deformation, e.g., near $t=T/2$ in case (ii), and near $t=T/4$ and $3T/4$ in case (iii). These deviations, though small, can accumulate over successive cycles due to the system’s sensitivity to initial conditions \cite{bonacci2023dynamics}, potentially giving rise to complex and even chaotic dynamics on longer timescales as we will show later.

To understand how such irreversibility emerges, we now analyze the change in orientation angle over one period, and identify the physical mechanisms that give rise to non-reciprocal dynamics, characterized by a difference in initial and final filament orientation $\delta \theta_0 = \theta_0(T) - \theta_0(0)$.

\subsection{Buckling and thermal fluctuations as drivers of irreversibility} \vspace{-0.2cm}

 To shed light on the respective roles of buckling deformations and thermal fluctuations in non-reciprocal filament dynamics we focus on numerical simulations. We consider weakly Brownian filaments with a persistence length to contour length ratio of $\ell_p/L=50$, large compared to experiments, such that purely thermal deformations are of negligible amplitude compared to buckling deformations. The  elastoviscous number is fixed at a large value well above the bucking threshold, $\bar{\mu}_{\text{m}} = 10^4 \gg \bar{\mu}_{\text{m}}^c$, to ensure that buckling can occur. Figure~\ref{fig:2} analyzes three representative trajectories of nearly straight filaments with initial orientations of $\theta_0^i=10^{\circ}$, $30^{\circ}$ and $90^{\circ}$ at a fixed dimensionless period $\rho=14$. It shows how both the orientation angle ($\theta_0$) and the filament deformation ($\|\boldsymbol{\Psi}\|$) evolve over time. To disentangle the respective roles of thermal fluctuations and filament deformations, we compare simulations for flexible Brownian and non-Brownian filaments to the case of a non-Brownian rigid rod, as well as a semi-analytical reduced-order model for a non-Brownian filament based on a truncated modal expansion of the shape vector (see SM \cite{Supplemental_Info} for details).

As can be seen from the successive filament shapes shown in the first row, and consistent with the observations of Fig.~\ref{fig:1}, the filament dynamics depend strongly on the initial orientation. We first focus on the Brownian simulations (orange symbols), and compare them to Jeffery dynamics of the non-Brownian rigid rod simulations (dashed blue curve). The data show that the evolution of the filament orientation (middle row) and deformation (bottom row) are strongly coupled, with significant deviations from Jeffery dynamics coinciding with buckling events. However, buckling alone does not guarantee non-reciprocal dynamics: all three examples undergo buckling, yet only the latter two fail to return to their initial orientation and shape by the end of the cycle. In the first case, with a low initial orientation angle near the flow axis, the filament spends the second half-cycle under extension and is able to fully relax back to a straight shape. This is not the case in the second and third examples: as the initial angle increases, the second half of the cycle becomes dominated by compression, leading to only partial stretching (second case) or no relaxation at all (third case, which ends in the compressional quadrant).

The comparison of Brownian and non-Brownian flexible simulations further highlights the role of thermal shape fluctuations in sustaining buckling and contributing to non-reciprocal dynamics. Non-Brownian simulations (purple square symbols) are performed by first initializing the filaments in a thermal bath to allow for weak initial shape fluctuations necessary to enable buckling, after which thermal noise is turned off and the flow is applied. The same process is applied for the semi-analytical reduced-order model (black curve). While the non-Brownian simulations and semi-analytical model agree closely with each other for all the three examples, their dynamics differ significantly from the Brownian case. Deformations are generally weaker and tend to be damped out over the course of the oscillation period for the non-Brownian system. In some cases (third example), buckling does not even occur in the absence of noise, even though it does in the Brownian version of the same scenario. When multiple periods are applied (not shown), non-Brownian systems always relax to a straight conformation that follows Jeffery dynamics \cite{TS2004}. 

A weakly nonlinear analysis performed on the reduced-order model provides analytical insight into these observations (see SM \cite{Supplemental_Info} for details). It shows that filament deformation results from the competition between two effects: an exponential growth term from the time-dependent viscous forcing, and a damping term from bending elasticity. The growth term depends on the accumulated time-dependent elastoviscous number and captures the combined effect of the flow’s unsteadiness and the filament’s tumbling motion. The damping term reflects the relaxation of bending modes, with the relaxation time of a mode with  wavenumber $q_k=(k+1/2)\pi/L$ given by \vspace{-0.0cm}
\begin{equation}
    \tau_k=(\zeta_\perp/B)q_k^{-4},
\end{equation}
where $\zeta_\perp$ is the transverse viscous drag coefficient. In our parameter regime, the longest relaxation time $\tau_1$ is typically comparable to the flow period $T$, meaning that any initial shape perturbations are damped away within a fraction of a cycle. Once these fluctuations are gone, there is no mechanism left to trigger the linear buckling instability that is the driver of large deformations (see bottom of Fig.~\ref{fig:2}(c)).

In summary, Fig.~\ref{fig:2} highlights that over one cycle, shape fluctuations, even minute, are essential to the irreversibility by continuously seeding buckling instabilities, thus enabling elastically driven non-reciprocal dynamics to occur. Fluctuations in orientation seem to play a subdominant role as significant deviations from Jeffery dynamics are generally associated with strong filament deformation. Over several cycles, the effect of orientational noise can, however, influence the long-term filament dynamics as we will discuss in the last section of the paper. 

\begin{figure}
  \centering
  \includegraphics[width=\columnwidth]{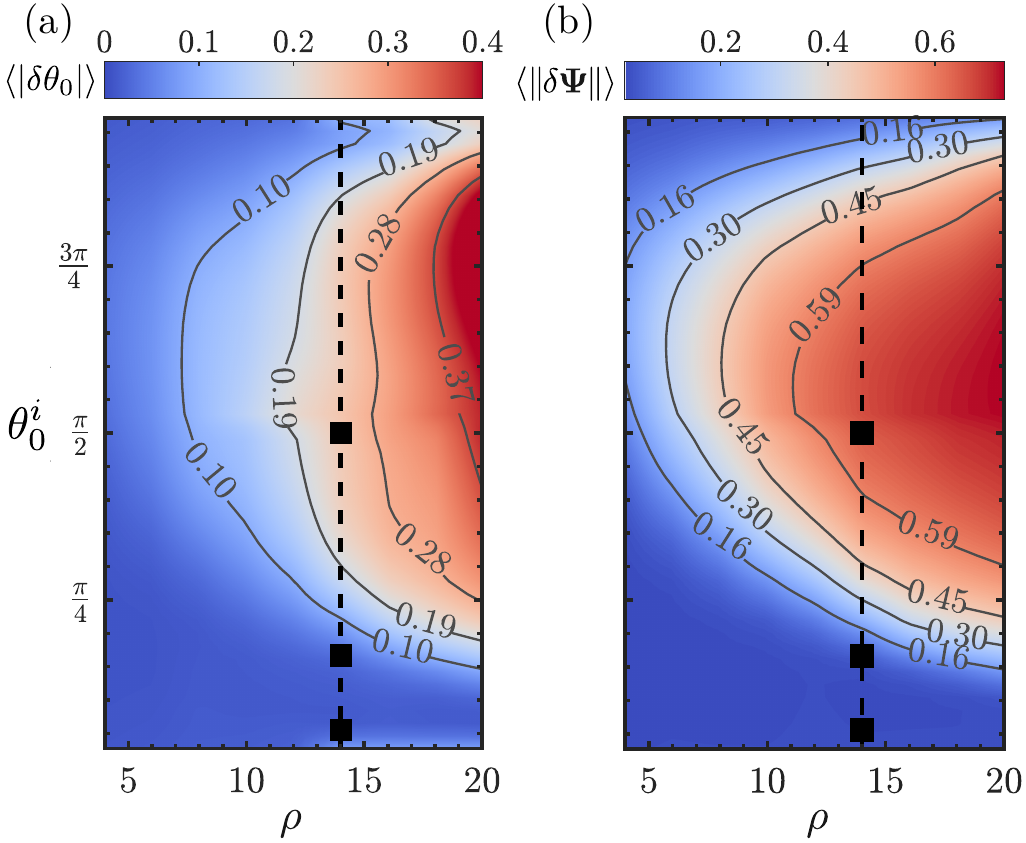} \vspace{-0.0cm}
    \caption{Ensemble-averaged stroboscopic changes in (a) mean filament orientation, $\langle | \delta \theta_0 | \rangle = \langle | \theta_0(T) - \theta_0(0) | \rangle$, and (b) filament deformation, $\langle \| \delta\boldsymbol{\Psi} \| \rangle = \langle \| \boldsymbol{\Psi}(T) - \boldsymbol{\Psi}(0) \| \rangle$, as functions of the dimensionless period $\rho$ and initial orientation $\theta_0^i$. Averages were computed from 50 simulations per $(\rho, \theta_0^i)$ pair, each with a different random seed. Simulations used $\ell_p/L = 50$ and $\bar{\mu}_m = 10^4$. Black squares indicate the three examples shown in Fig.~\ref{fig:2}.} \vspace{-0.cm}
  \label{fig:3}
\end{figure}

We synthesize these findings in Fig.~\ref{fig:3}, where, starting from a straight filament, we plot the ensemble-averaged stroboscopic differences after one period in the filament's mean orientation  $\langle | \delta \theta_0  | \rangle =  \langle | \theta_0(T) - \theta_0(0) | \rangle$  and deformation $\langle \|  \delta\mathbf{\Psi}  \| \rangle = \langle \| \mathbf{\Psi}(T) - \mathbf{\Psi}(0) \| \rangle$ as functions of ($\rho$, $\theta_0^i$), obtained from 
50 numerical simulations of Brownian filaments with identical conditions but different random seeds. The structure of the phase diagrams can be explained as follows. For small $\rho$, the time spent under compression is short, so buckling is rare and dynamics are effectively reversible regardless of initial angle. As $\rho$ increases, the probability of buckling events becomes significant, and these events can result in stroboscopic differences depending on the initial filament orientation. If the initial angle is very small corresponding to a mean orientation close to the negative flow direction ($\theta_0^i\gtrsim 0$),  the filament spends most of its time in the compressive quadrant of the flow for $t<T/2$, and in the extensile quadrant for $t>T/2$ (see cases (i), (ii) and (iii) in Fig.~\ref{fig:1}(c), as well as Fig.~\ref{fig:2}(a)). The long stretch under extension allows the filament to straighten out before the end of the cycle, leading to quasi-reversible stroboscopic dynamics even if buckling occurs mid-cycle. For larger initial angles ($\theta_0^i\gtrsim \pi/2$), the filament spends most of the first half-period in the extensional quadrant until it aligns near the positive flow direction at $t = T/2$, and undergoes strong compression in the second half of the cycle with little to no time for relaxation (see cases (iv) and (v) in Fig.~\ref{fig:1}(c), as well as Fig.~\ref{fig:2}(b,c)). This leads to large stroboscopic differences in orientation and shape, signaling irreversible dynamics. For even larger initial angles ($\theta_0^i\lesssim \pi$), filaments mostly remain trapped very close to the flow direction and compression is not strong enough in the second half of the cycle to induce significant buckling. For these cases, stroboscopic differences are again small and the dynamics can be considered to be reversible.  Irrespective of $\theta_0$, filaments only very rarely cross the flow direction due to the strong slowdown of the angular velocity close to the flow axis. As a consequence, filament dynamics are mostly restricted to one or possibly two flow quadrants, as in all the examples described here. With increasing $\rho$,  the region of dynamical irreversibility widens, as ever smaller (or larger) initial angles can map to orientations prone to buckling in the second half cycle. 

In conclusion, the structure of the stroboscopic irreversibility maps of Fig.~\ref{fig:3} reflects the interplay between buckling and relaxation under time-dependent viscous stresses, with a minor contribution from rotational diffusion. Strong deformations associated with buckling result in deviations from the Jeffrey dynamics and are the primary driver of irreversibility. Although Brownian shape fluctuations alone generate only weak deformations in the present conditions, they are crucial for triggering and sustaining buckling.

\subsection{Long-time dynamics: coexisting quasi-reversible attractors and intermittent transitions}
\label{subsec:longtime}

\begin{figure*}[t]
  \centering
  \includegraphics[width=\textwidth]{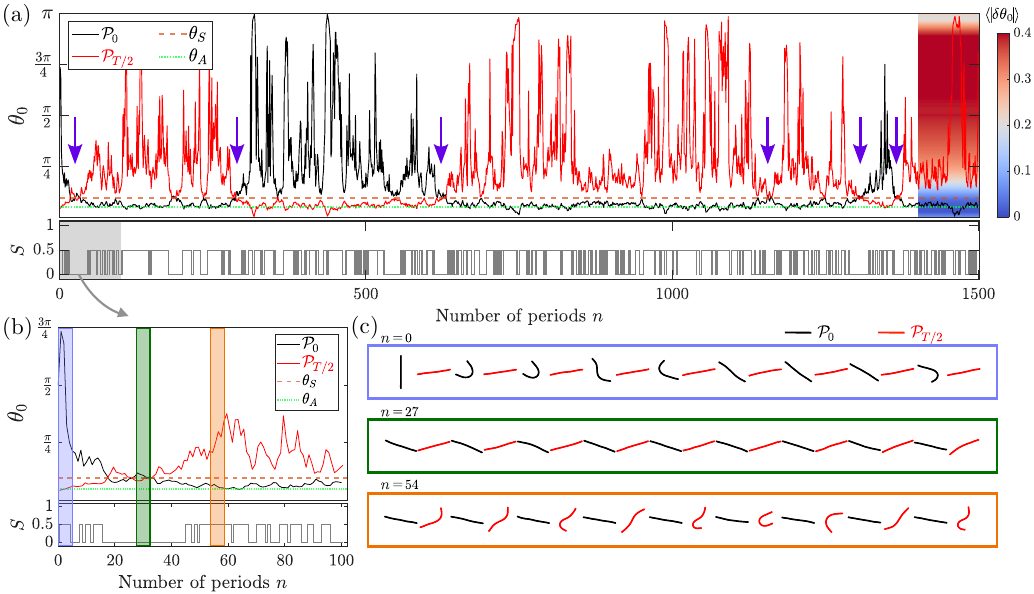} \vspace{-0.1cm}
  \caption{(a) Top panel: Evolution of the two stroboscopic sequences $\mathcal{P}_0(n)$ (black) and $\mathcal{P}_{T/2}(n)$ (red) over 1500 periods in a Brownian simulation with $\rho = 20$ and $\bar{\mu}_m = 1 \times 10^4$. The orange dashed line indicates the switching angle $\theta_S$ between the two time series, and the green dotted line marks the mean attracting angle $\theta_A$ for stable dynamics. The phase chart from Fig.~\ref{fig:3}(a), showing the ensemble-averaged stroboscopic change $\langle | \delta \theta_0 | \rangle$, is overlaid on the right. Purple arrows highlight crossings between the two sequences. Bottom panel: Time evolution of the order parameter $S$ that characterizes filament deformation states (see main text for definition). (b) Zoom-in on the first 100 periods from (a), highlighting three distinct dynamical regimes using blue, green, and orange interrogation windows. (c)~Filament morphologies corresponding to the three windows in (b). The top and bottom rows depict the two attracting states, where one sequence is quasi-reversible and the other chaotic, resulting in $S = 0.5$. The middle row shows a switching event, in which both sequences are reversible and the filament adopts symmetric straight conformations, yielding $S = 0$. Also see Supplemental Videos 1 to 3 \cite{Supplemental_Info} for movies showing the dynamics.}\vspace{-0.0cm}
  \label{fig:4}
\end{figure*}

Having explored buckling and irreversibility over a single period, we now turn to the long-time statistical behavior of a single filament. Experimentally tracking filaments over hundreds of cycles is difficult because Brownian motion often causes them to drift out of the focal plane. We therefore continue to focus on numerical simulations, with the persistence length ratio $\ell_p/L=50$ and elastoviscous number $\bar{\mu}_{\text{m}} = 10^4$, as in the previous section. As shown earlier, these values minimize thermal shape fluctuations while still allowing them to induce buckling.


To investigate orientational dynamics over time, we monitor the filament's orientation angle $\theta_0$ at the start of each period. This gives a discrete-time stroboscopic sequence, denoted $\mathcal{P}_0(n)= \theta_0(n T)$, where $n$ is a non-negative integer. This time series is equivalent to taking a Poincar\'e section of the $(\theta_0,\dot{\theta}_0)$ phase space at $\dot{\theta}_0=0$. Owing to the system’s time-glide symmetry, we also define a complementary stroboscopic sequence at the half-period as $\mathcal{P}_{T/2}(n)=\pi- \theta_0(nT+T/2)$. These two sequences are statistically identical but correlated, since each depends on the filament’s conformational history over previous cycles.

Figure~\ref{fig:4}(a) shows both $\mathcal{P}_0$ and $\mathcal{P}_{T/2}$ (black and red curves, respectively) over 1500 cycles for a simulation with $\rho = 20$. Both sequences exhibit intermittent dynamics characterized by prolonged stretches (hundreds of periods) of stable quasi-reversible behavior, where the stroboscopic orientation angle hovers around a low constant value, alternating with equally long bursts of chaotic irreversible activity, where the angle fluctuates wildly. 

As anticipated, these dynamics correlate with the 
behavior characterized previously over one cycle: quasi-reversible intervals occur when the filament is aligned close to the flow axis, while chaotic fluctuations involve larger orientation angles. This is confirmed by overlaying the phase chart of $\langle | \delta \theta_0 | \rangle$ from Fig.~\ref{fig:3}(a) on the right side of Fig.~\ref{fig:4}(a), which shows that stable dynamics occur for angles within the blue reversible range, while chaotic dynamics are observed within the red irreversible range. Intermittency in the system is typified by the occasional switching of each stroboscopic sequence between chaotic and stable dynamics.


Strikingly, the two stroboscopic sequences $\mathcal{P}_0$ and $\mathcal{P}_{T/2}$ exhibit alternating behavior: when one is stable, the other is chaotic. They occasionally cross,  as indicated by vertical purple arrows in Fig.~\ref{fig:4}(a), signaling a transition where one time series becomes irreversible and the other becomes reversible. To illustrate this exchange, Fig.~\ref{fig:4}(b) zooms in on the first 100 oscillation periods, revealing that the crossover is gradual rather than abrupt, extending over several periods and featuring short-lived intersections. The figure also highlights three interrogation windows whose filament conformations are shown in Fig.~\ref{fig:4}(c) (also see Supplemental Videos 1 to 3 for movies showing the dynamics \cite{Supplemental_Info}). In the early blue window ($n \lesssim 9$), $\mathcal{P}_{T/2}$ exhibits stable, reversible dynamics both for orientation and shape. In contrast, during the orange interval ($54 \lesssim n \lesssim 63$), $\mathcal{P}_0$ is stable and reversible. The green window ($27 \lesssim n \lesssim 36$) captures a crossover, where both sequences appear reversible. This is corroborated by the nearly symmetric and reversible conformations shown in Fig.~\ref{fig:4}(c). Together, these observations point to the existence of attracting states featuring an absorbing, straight conformation closely aligned with the negative flow direction on each period, 
alternating with a deformed state on the half-period. Owing to the time-glide symmetry of the system, two such states coexist with a half-period phase shift. The filament spontaneously selects one of these states but intermittently switches between them due to fluctuations, producing alternating stretches of stroboscopic quasi-reversibility and apparent chaos.

We now examine systematically how the temporal evolution of the stroboscopic sequences relates to filament conformational dynamics. To quantify the filament’s deformation state, we define a binary order parameter $\tilde{S}$ by assigning $\tilde{S} = 1$ if the filament is deformed at the start of a half period, and $\tilde{S} = 0$ if it is straight. This yields a time series sampled every half period. A threshold of $ \| \mathbf{\Psi} \| = 0.05$ is used to distinguish between straight ($\| \mathbf{\Psi} \| < 0.05$) and deformed ($\| \mathbf{\Psi} \| > 0.05$) filaments \cite{bonacci2023dynamics}. From the $\tilde{S}$ time series, we then compute a moving average over a full period to define $S$, which takes on three discrete values:
\begin{itemize}
	\item $S=0$: the filament is undeformed at the end of both half cycles (deformation sequence blue--blue in the nomenclature of Fig.~\ref{fig:1}(a)); 
	\item $S=0.5$: the filament is deformed during one half cycle and undeformed during the other (sequences blue--red and red--blue);
	\item $S=1$: the filament is deformed at the end of both half cycles (sequence red--red). 
\end{itemize}
The bottom panels of Fig.~\ref{fig:4}(a,b) show the evolution of $S$, revealing a strong correlation between the stroboscopic sequences and conformational state. In particular, $S = 0$ consistently coincides with crossings of the two sequences (green window in Fig.~\ref{fig:4}(b,c)), indicating symmetric conformational dynamics. This can correspond to pure tumbling without deformation or to a buckling event followed by symmetric stretching in both half cycles, as seen in case (iii) of Fig.~\ref{fig:1}(c). For the large $\bar{\mu}_m$ and $\rho$ values considered here, buckling is observed. Thus, in practice, $S = 0$ typically reflects alternating buckling and stretching in the two half cycles of the period. $S=0.5$ is the most common case [Fig.~\ref{fig:4}(a)], where one stroboscopic sequence is stable and the other chaotic.  $S=1$ would correspond to a situation in which the filament spends enough time under compression in a first half period to get deformed but not enough time to relax in the second half period. Because of the symmetry of the Jeffery dynamics underlying the filament dynamics, this situation is rarely observed. 

\begin{figure}[t]
\centering
\includegraphics[width=\columnwidth]{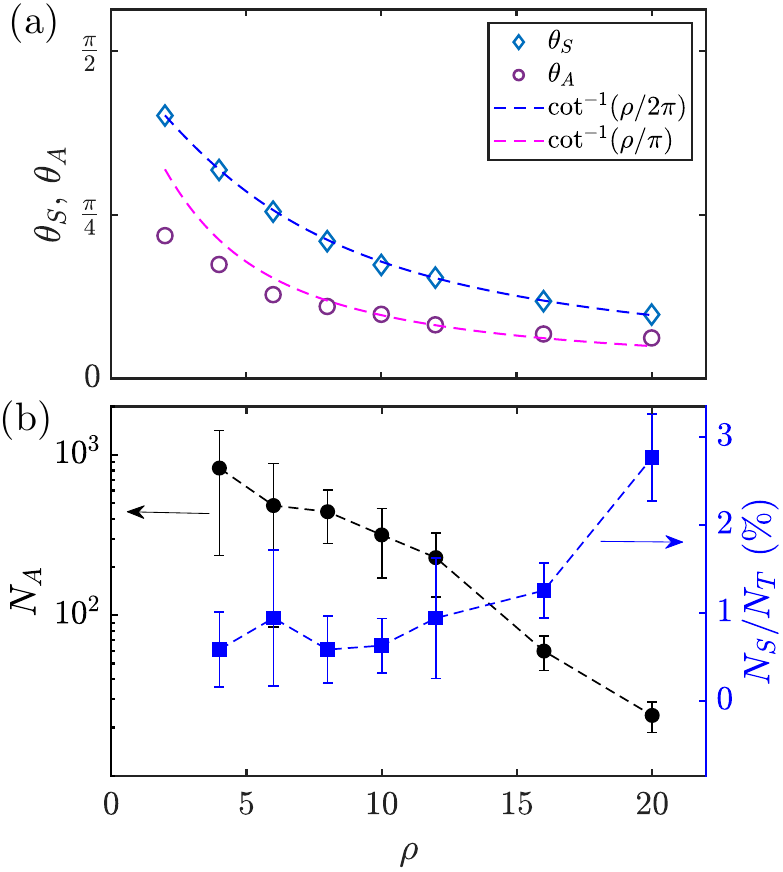}
    \caption{(a) Switching angle $\theta_S$ marking the transition between stable and chaotic dynamics, and attracting angle $\theta_A$ during stable dynamics, shown as functions of the dimensionless period $\rho$. Symbols represent simulation results; dashed lines indicate theoretical predictions. (b) Mean residence time $N_A$ in a stable attracting state, and the normalized switching frequency $N_S/N_T$, where $N_S$ is the number of transitions between stable and chaotic dynamics and $N_T = 2500$ is the total number of periods. Error bars indicate standard deviations. Averages were computed from 12 long Brownian simulations with $\bar{\mu}_m = 1 \times 10^4$ and different initial orientations. All results correspond to Brownian filaments with $\ell_p/L = 50$.}
\label{fig:5}
\end{figure}

Crossings of $\mathcal{P}_0$ and $\mathcal{P}_{T/2}$ mark symmetry points where both states temporarily converge to reversible behavior, and correspond to filament orientations that divide the dynamics into distinct attractor basins. In orientational space, such events are nearly symmetric about $\theta = \pi/2$, implying $\theta(0) = \pi - \theta(T/2)$. For a given $\rho$, this condition defines a unique orientation $\theta_S$ at which the stroboscopic sequences intersect and the dynamics become reversible in phase space. Based on the Jeffery dynamics of Eq.~(\ref{eq:Jeffery}), we obtain this special angle as $\theta_S = \cot^{-1}(\rho / 2\pi)$. Note that case (iii) of Fig. \ref{fig:1} corresponds to such a situation. This prediction, shown as an orange dashed line in Fig.~\ref{fig:4}(a,b), aligns well with the observed crossing points of the stroboscopic sequences, confirming that $\theta_S$ serves as the boundary between the system’s two equiprobable attractors. To further test this prediction, we performed simulations across a range of $\rho$ values using an ensemble of initial filament configurations to numerically identify the angle $\theta_S$ at which the sequences intersect. The results, shown in Fig.~\ref{fig:5}(a), demonstrate excellent agreement with the theoretical prediction.


During stretches of quasi-reversible dynamics, the stable stroboscopic sequence hovers around a preferred filament orientation, exhibiting only minimal fluctuations. We denote this orientation by $\theta_A$, which is identified numerically as the peak in the stroboscopic angular distribution and is shown as a green dotted line in Fig.~\ref{fig:4}(a,b). As the period $\rho$ increases, the filament spends more time aligning with the flow during extension, leading to a gradual decrease in $\theta_A$. This trend is shown in Fig.~\ref{fig:5}(a), where $\theta_A$ is plotted as a function of $\rho$. For sufficiently large $\rho$, we find empirically that $\theta_A \approx \cot^{-1}(\rho/\pi)$, indicated by the pink dashed line. For short periods, the distinction between reversible and irreversible behavior is less clear, and $\theta_A$ is no longer well-defined. A physical interpretation of the empirical relation can be gleaned from Jeffery dynamics: a rigid rod with initial orientation $\theta_0^i = \cot^{-1}(\rho/\pi)$ follows a trajectory in $(\theta_0, \dot{\theta}_0)$ phase space that becomes tangent to $\theta_0 = \pi/2$ at $T/2$. This corresponds to an initial orientation comprised between case (i) and (ii) shown in Fig. \ref{fig:1}.  This means the filament experiences pure compression during the first half-period and pure extension during the second. As a result, $\theta_A$ corresponds to the orientation most likely to induce buckling followed by full relaxation, yielding an expected deformation parameter value of $S = 0.5$ (see Fig.~\ref{fig:4}(b)).

As shown in the time series of Fig.~\ref{fig:4}(a), switching between the two attracting states occurs stochastically on long time scales, typically spanning hundreds of oscillation periods. To quantify this intermittency, we define two measures: (i) the mean residence time $N_A$, i.e., the average number of periods spent in a given attracting state, and (ii) the mean switching frequency $N_S/N_T$, where $N_S$ is the number of switching events and $N_T$ is the total number of periods in the time series. For a Poisson process, these quantities are inversely related. Figure~\ref{fig:5}(b) shows both $N_A$ and $N_S/N_T$ as functions of $\rho$. At low $\rho$, filament deformations are negligible (see Fig.~\ref{fig:3}), and the system remains stroboscopically trapped near its initial orientation, exhibiting only weak fluctuations due to rotational diffusion. Filaments initialized near $\theta_A$ are therefore likely to remain in the same attracting state for long durations, resulting in large $N_A$ and infrequent switching. As $\rho$ increases, buckling events become more frequent, enhancing stochasticity in the conformational dynamics [Fig.~\ref{fig:5}(b)]. Simultaneously, the angular separation between the attracting angle $\theta_A$ and the switching angle $\theta_S$ decreases [Fig.~\ref{fig:5}(a)], lowering the threshold for transitions. These combined effects increase the probability of noise-induced switching, as reflected in the decreasing trend of $N_A$ and the rising trend of $N_S/N_T$ with increasing $\rho$ in Fig.~\ref{fig:5}(b).

\begin{figure}
\centering
\includegraphics[width=\columnwidth]{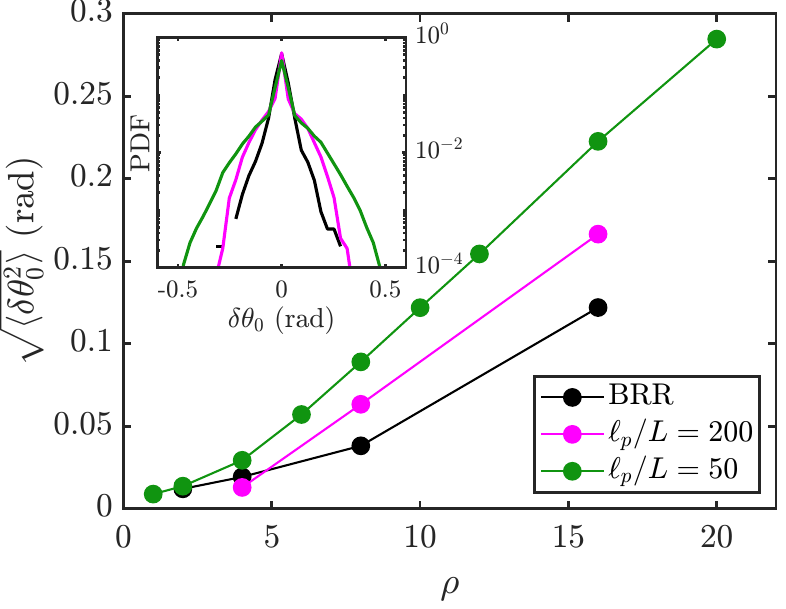}\vspace{-0.0cm}
\caption{Standard deviation $\sqrt{\vphantom{\delta}\smash{\langle \delta\theta_0^2\rangle}}$ of the stroboscopic change in the mean orientation angle over one oscillation period, plotted as a function of the dimensionless frequency $\rho$ for a Brownian rigid rod (BRR) and Brownian filaments with two values of $\ell_p/L$. Inset: Probability density functions of $\delta\theta_0$ for the same three cases at $\rho = 8$. For the BRR model, we used a P\'eclet number of $\mathrm{Pe} = \dot{\gamma}_m/d_r = \bar{\mu}_m c \ell_p/L /(24 \log(2/\varepsilon))$, where $d_r$ is the rotational diffusivity of the rigid rod. This choice ensures identical flow conditions between the flexible filament and the BRR to facilitate their comparison \cite{chakrabarti2021signatures}. }
\label{fig:6}
\end{figure}

Taken together, these results indicate that the effective fluctuations in the filament’s orientational dynamics grow with increasing oscillation period. These fluctuations arise from two sources: thermal orientational noise and buckling-induced deviations from deterministic Jeffery orbits, shape fluctuations being limited to small amplitudes. To disentangle the different contributions, we analyze the standard deviation $\smash{\sqrt{\langle \delta\theta_0^2\rangle}}$ of the stroboscopic angular displacement as a measure of the accumulated orientational noise over one oscillation period. Figure~\ref{fig:6} shows this standard deviation as a function of $\rho$ for a Brownian rigid rod and two Brownian flexible filaments with different $\ell_p/L$ ratios.
In all cases, the effective noise, quantified by $\sqrt{\langle \delta\theta_0^2\rangle}$, increases monotonically with $\rho$. For the rigid rod, this growth is purely due to thermal rotational diffusion acting over longer time intervals as $\rho$ increases, allowing the orientation to deviate further from its initial value. When filament flexibility is introduced, the noise is significantly amplified relative to the rigid rod case. This enhancement arises from the additional stochasticity associated with buckling, which itself is triggered by thermal shape fluctuations and becomes more important with decreasing $\ell_p/L$. 
For even smaller $\ell_p/L$ than investigated here, closer to experimental conditions, yet more complex dynamics may arise as filaments become more likely to explore more than two quadrants during a half-period, as shape fluctuations facilitate tumbling of the filament across the flow direction.

At low $\rho$, the magnitude of fluctuations is nearly identical across all cases, as buckling remains negligible and thermal orientational diffusion dominates. {Further insight into the relative contributions of thermal noise and buckling is provided by the corresponding probability density functions of $\delta\theta_0$ shown in the inset of Fig.~\ref{fig:6} for the case $\rho=8$. All three distributions (rigid and flexible) coincide at small angular displacements, at which orientational diffusion plays a significant role: this corresponds to the quasi-reversible regime where the stroboscopic orientation hovers around $\theta_A$. Larger angular displacements, however, are much more probable in the case of flexible filaments and correspond mostly to buckling events in the chaotic regime, an effect that is enhanced with decreasing $\ell_p/L$.} Altogether, these results highlight how the coupling of thermal fluctuations and buckling effectively amplifies orientational noise, thus facilitating stochastic transitions between the system's two basins of attraction.

\section{Concluding remarks} We have analyzed the dynamics of semiflexible filaments in oscillatory shear flow, a canonical nonlinear, stochastic Floquet system exhibiting time-glide symmetry under weak thermal fluctuations. Although thermal noise is weak, it plays a crucial role by triggering elastoviscous buckling, which serves as the primary stochastic mechanism driving departures from deterministic rigid-rod behavior.
Our focus has been on the emergence of irreversibility as the oscillation period, $\rho$, increases. We identified two distinct regimes. In fast-oscillating flows, buckling has insufficient time to develop, and filaments undergo quasi-reversible oscillations while remaining nearly straight, regardless of their orientation. At higher $\rho$, when the period of oscillation is slower, buckling becomes more likely and leads to irreversibility.

Despite the onset of buckling, we uncovered the existence of attracting states in which the filament remains nearly straight and aligned with the flow at the start of each period, alternating with a deformed conformation at mid-cycle. Due to the system’s time-glide symmetry, two such states coexist, out of phase by half a period. Filaments spontaneously select one, but intermittently switch between them due to noise. This results in a stroboscopic dynamic characterized by alternating stretches of quasi-reversibility and chaotic behavior, with the typical duration of each stretch decreasing as $\rho$ increases.

Whether these single-filament dynamics have observable signatures on large scales at the suspension level remains an open question. Under steady shear, buckling instabilities enhance shear thinning and generate normal stress differences \cite{BS2001,chakrabarti2021signatures}. Similar effects may emerge under oscillatory flow. A particularly intriguing possibility is that hydrodynamic interactions could synchronize reversible and chaotic phases among nearby filaments, as synchronization is prone to happen in other systems of driven or active hydrodynamically coupled elastic structures \cite{CS2019,TL2022,CFS2022}. If so, this could lead to intermittency in the macroscopic rheological behavior, suggesting new routes for designing and controlling responsive complex fluids with unconventional flow properties.

\begin{acknowledgments} 

The authors thank Muhittin Mungan for a careful reading of the manuscript and thoughtful comments, and Rama Govindarajan for discussions. B.C. acknowledges the support of the Department of Atomic Energy, Government of India, under project no. RTI4001. F.B. acknowledges the  European Union – NextGenerationEU under
the Italian Ministry of University and Research (MUR) National Innovation
Ecosystem grant ECS00000041 -- VITALITY -- CUP J97G22000170005.
F.B., A.L. and O.d.R. acknowledge funding from the European Research
Council (ERC) Consolidator Grant PaDyFlow (Agreement 682367) and the support of Institut Pierre-Gilles de Gennes (Equipement d’Excellence, “Investissements d’avenir,” program ANR-10- EQPX-34). O.d.R acknowledges the funding from the French Agence Nationale de la Recherche (ANR-21-CE13-0048 and ANR-22-CE30-0024) and is a member of GDR 2108 Quantitative Approaches of Life. D.S. acknowledges funding from National Science Foundation Grants No. 2153520 and 2327243.

\end{acknowledgments} 

\section*{Author contributions}

All authors designed the research. F.B. performed the experiments and developed the reduced-order model. B.C. developed the numerical model and simulation software. F.B. and B.C. performed the simulations, analyzed data and prepared figures. All authors interpreted the data and wrote the manuscript.

\appendix
	
\section{Experimental methods\label{appendixA}}

We use F-actin filaments as a model system for Brownian semiflexible polymers. The filaments are stabilized and fluorescently labeled with phalloidin. Their contour lengths $L$ range from 5 to 25 µm, with diameter approximately 8 nm and persistence length $\ell_p \approx 17\,$µm. This corresponds to a bending modulus of $B = k_B T \,\ell_p \approx 7 \times 10^{-26} \,\mathrm{N \cdot m^2}$. Further details on filament synthesis and mechanical characterization are available in our previous studies \cite{LCDLD2018, bonacci2023dynamics}. A schematic of the oscillatory flow setup is provided in Fig.~S3 of the SM \cite{Supplemental_Info}. We generate a time-periodic two-dimensional linear shear flow using a vertical Hele–Shaw channel with a rectangular cross-section, designed to have a height-to-width ratio of approximately 3.33. This geometry produces an almost plug-like flow profile in the vertical direction and a Poiseuille-like profile in the observation plane, where filaments are imaged at roughly 100 µm from the bottom coverslip. Experiments focus on filaments positioned midway between the sidewall and channel center, ensuring the flow gradient is approximately uniform over the filament length \cite{bonacci2023dynamics, LCDLD2018}. The oscillatory flow is driven by a pair of pressure controllers (Fluigent Lineup Flow EZ\textsuperscript{TM}) that impose a sinusoidal pressure drop $\delta p \sin(2\pi t / T)$ with periods $T$ ranging from 1 to 10$\,$s. In all experiments, the filament Reynolds numbers are very low, between $10^{-5}$ and $10^{-4}$, and the maximum shear rates achievable range from $\dot{\gamma}_m = 0.4$ to $3.7\,\mathrm{s}^{-1}$. These conditions are attained by finely tuning the pressure amplitude $\delta p$ at the channel inlets, which remains below a few millibars. The low pressure drops are facilitated by incorporating high-resistance variable length tubing between the inlets and the filament suspension \cite{bonacci2023dynamics}. Given the relatively low oscillation frequencies, viscous resistance dominates over inertial transients, ensuring the flow remains laminar and Poiseuille-like throughout each oscillation cycle \cite{bonacci2023dynamics}. This guarantees that the instantaneous shear rate experienced by the filaments is in phase with the imposed pressure, such that $\dot{\gamma}(t) = \dot{\gamma}_m \sin(2\pi t / T)$. Representative shear rate data and sketches of the flow profiles at various times are shown in Fig.~\ref{fig:1}(a). Additional details on the experimental setup, image acquisition, and analysis can be found in our previous work \cite{bonacci2023dynamics}. \vspace{0.cm}

\section{Model governing equations and computational methods\label{appendixB}}

We model slender filaments as one-dimensional inextensible space curves, represented by their centerline position $\boldsymbol{r}(s,t)$, parametrized by arclength $s \in [0,L]$. Hydrodynamics are captured via local slender body theory for Stokes flow, where viscous anisotropic drag balances bending elasticity, internal tension, and Brownian forces.
All variables are nondimensionalized by scaling lengths by the filament contour length $L$, time by the characteristic bending relaxation time $\tau_r = 8 \pi \mu L^4 / B$, Brownian forces by $\smash{\sqrt{\vphantom{\ell}\smash{L/\ell_p}}\,B/L^2}$, and elastic forces by $B / L^2$. The filament dynamics are governed by the following equation of motion:
\begin{equation} \label{eq:SBE}
	\partial_{t} \boldsymbol{r}(s, t)=\bar{\mu}_{m} \boldsymbol{u}^{\infty}(s,t) + \boldsymbol{\mathcal{M}} \cdot\big[-\boldsymbol{r}_{s s s s} + \left(\Lambda \boldsymbol{r}_{s}\right)_{s}+\sqrt{\vphantom{\ell}\smash{L/\ell_p}}\, \boldsymbol{\zeta}\big],
\end{equation}
where $\boldsymbol{u}^\infty(s,t)$ is the imposed background oscillatory shear flow, $\boldsymbol{\mathcal{M}} = \boldsymbol{I} + \boldsymbol{r}_s \boldsymbol{r}_s$ is the local anisotropic mobility matrix, $\Lambda(s)$ is a line tension that acts as a Lagrange multiplier to impose the inextensibility constraint $\boldsymbol{r}_s \cdot \boldsymbol{r}_s = 1$, and $\boldsymbol{\zeta}(s,t)$ is a Gaussian white noise vector with zero mean and unit variance. The filament dynamics depend on three key dimensionless parameters. The first one is the maximum elastoviscous number, which quantifies the strength of hydrodynamic forcing and is given by: $\bar{\mu}_{m} = 8 \pi \mu \dot{\gamma}_{m} L^{4}/B c$, where $c = -\log(e\varepsilon^2)$ is the slenderness parameter with $\varepsilon=a/L\ll1$ representing the filament's aspect ratio. The second number is the dimensionless period $\rho = \dot{\gamma}_{m} T$, which controls the time scale of oscillations. This parameter enters the time-periodic external flow, defined as $\boldsymbol{u}^\infty(s,t) = \left[\sin\left(2 \pi \bar{\mu}_{\text{m}} c t/\rho\right) y,0,0\right]$. The third dimensionless number is $L/\ell_p$, which characterizes the magnitude of thermal shape fluctuations. Further details on the numerical methods employed for solving these equations can be found in \cite{LCDLD2018, chakrabarti2021signatures, TS2004}.

\bibliography{references}

\end{document}